\documentclass[11pt]{article} 

\usepackage{graphicx}
\usepackage{booktabs}
\usepackage[T1]{fontenc}
\usepackage{lmodern}
\usepackage[flushleft]{threeparttable}
\usepackage[hidelinks]{hyperref} %
\hypersetup{hidelinks}
\usepackage{multirow}
\usepackage{longtable}
\usepackage[margin=3cm]{geometry} %
\usepackage{amsmath}
\usepackage{authblk}
\usepackage{caption}

\DeclareMathOperator\erf{erf}

\title{Investigation and rectification of NIDS datasets and 
standardized feature set derivation for network attack detection with graph neural networks}
\author[1]{Anton Raskovalov}
\author[1,*]{Nikita Gabdullin}
\author[1]{Vasily Dolmatov}
\affil[1]{Joint Stock "Research and production company "Kryptonite" \authorcr
E-mail: a.raskovalov@kryptonite.ru, n.gabdullin@kryptonite.ru, v.dolmatov@kryptonite.ru}
\affil[*]{corresponding author}
\date{}
\begin{document}

    \captionsetup[table]{labelformat={default},labelsep=period,name={Table}}

    \maketitle

    \begin{abstract}
        Network Intrusion and Detection Systems (NIDS) are essential for malicious traffic and cyberattack detection in modern networks. 
        Artificial intelligence-based NIDS are powerful tools that can learn complex data correlations for accurate attack prediction. 
        Graph Neural Networks (GNNs) provide an opportunity to analyze network topology along with flow features which makes them 
        particularly suitable for NIDS applications. However, successful application of such tools requires large amounts of carefully 
        collected and labeled data for training and testing. In this paper we inspect different versions of ToN-IoT dataset and point out 
        inconsistencies in some versions. We filter the full version of ToN-IoT and present a new version labeled ToN-IoT-R. To ensure 
        generalization we propose a new standardized and compact set of flow features which are derived solely from NetFlowv5-compatible data. We 
        separate numeric data and flags into different categories and propose a new dataset-agnostic normalization approach for numeric 
        features. This allows us to preserve meaning of flow flags and we propose to conduct targeted analysis based on, for instance, 
        network protocols. For flow classification we use E-GraphSage algorithm with modified node initialization technique that allows us 
        to add node degree to node features. We achieve high classification accuracy on ToN-IoT-R and compare it with previously published 
        results for ToN-IoT, NF-ToN-IoT, and NF-ToN-IoT-v2. We highlight the importance of careful data collection and labeling and 
        appropriate data preprocessing choice and conclude that the proposed set of features is more applicable for real NIDS due to
        being less demanding to traffic monitoring equipment while preserving high flow classification accuracy.   
    \end{abstract}

    
    \section{Introduction}\
\label{introduction}

Information security has been one of the most important topics in Computer Science in 21\textsuperscript{st} century due to its importance for individuals, 
industries, and nations alike. In addition to traditional computer networks, IoT networks that include various portable, smart, and
edge devices also become a lucrative target for cyberattacks which are getting more and more sophisticated and rapidly grow in number. 
This makes the task of traffic monitoring performed by network intrusion detection systems (NIDS) one of paramount importance. 
Conventional methods focus on identifying characteristic features of specific attacks and use signature matching techniques for attack 
detection. However, such methods require constant updating due to everchanging pool of cyber threats. Furthermore, they struggle to 
identify previously unseen attacks, which is also known as zero-day attack problem.

Machine learning (ML) based methods have been steadily gaining popularity due to their ability to accurately capture complex data 
patterns and detect abnormalities. Recently graph neural networks (GNNs) have shown promising results compared to conventional ML techniques. 
Whereas the latter only work with tabular data, GNNs can capture network topology in addition to features of network flows providing an 
opportunity to seamlessly combine both in a single analysis framework. Moreover, graph representation network data does not require any 
additional manipulation when sources and destinations are represented as nodes and network flows are represented as graph edges.

Currently available NIDS datasets for GNN training and testing vary significantly in their size, attack sets, and flow feature sets \cite{NFp}. 
It was recently proposed to use a set of standardized features to ensure the applicability of GNNs to different networks \cite{NFp,NFv2p} with 
the optimal feature set still being a subject of investigation. However, there are some inconsistencies in different dataset versions which 
might affect GNN training and result interpretation.

To address this issue, we analyze different versions of ToN-IoT dataset \cite{TNDp}. We propose a new set of features along with a 
preprocessing normalization technique that preserves the meaning of the original flow features. For attack classification based on 
flow features we apply E-GraphSage \cite{EGSp} GNN in our experiments and obtain similar accuracy as compared with current leader \cite{NFv2p} 
on ToN-IoT dataset with a more compact set of features. We show that due to the interpretability of the proposed feature set 
additional analysis logic can be constructed allowing TCP and UDP flows to be analyzed separately. The proposed feature 
set is highly versatile and compatible with data obtained from conventional network monitoring devices.

The rest of the paper is organized as follows: Section~\ref{NIDS-datasets} discusses different versions of the ToN-IoT datasets, different flow feature sets
and provides summary of the dataset and feature set we propose; Section~\ref{GNN} discusses E-GraphSage and graph initialization methods, 
Section~\ref{Res-Disc} shows experimental results and Section~\ref{conclusions} concludes the paper.

\section{NIDS datasets}
\label{NIDS-datasets}

\subsection{ToN-IoT and its variations}
\label{ToN-IoT-vers}

Representative datasets are necessary for successful training of ML models. There are several publicly available NIDS datasets 
including BoT-IoT \cite{Botp}, ToN-IoT \cite{TNDp}, CSE-CIC-IDS \cite{CSEp}, and others. A comprehensive dataset comparison can be found in 
\cite{SURVp, NFp}. However, working with some of the original datasets is challenging due to their poor balance, limited number of 
represented attack classes, and inconsistent feature sets. For instance, BoT-IoT consists of 99.99\% 
attacks with only 4 attack classes, and CSE-CIC-IDS and UNSW-NB15 \cite{UNSWp} are predominantly benign. We focus on ToN-IoT due to its 
sufficiently large size and relatively good balance among 9 attack classes.

Table~\ref{tab:1} summarizes the details of four available versions of ToN-IoT. In the original paper the authors provide the full version 
along with a significantly smaller test-train subset, both of which have 45 flow features in dataset. NF-ToN-IoT is proposed by 
Sarhan \textit{et al.} \cite{NFp} with only 14 features used to describe the flows. NF-ToN-IoT-v2 is proposed by the same authors and it contains 
45 flow features that are different from the original ToN-IoT ones \cite{NFv2p}. NF-ToN-IoT and NF-ToN-IoT-v2 versions have been obtained 
by applying different feature extraction tools to original pcap files. 

However, there are some issues with ToN-IoT-based datasets. For instance, in NF-ToN-IoT a number of flows are duplicated and labeled 
as corresponding to different attacks. In all datasets there are flows where traffic directed outside the test network is recorded 
as some form of attack, but such traffic should not be considered in the analysis and training. There are also flows labeled as 
attacks that contain normal data exchange with network router and DNS resolver (IP: 192.168.1.1). To exclude such cases, we filter 
the full version of ToN-IoT (“Processed Network Dataset” files in \cite{TNDp}) and obtain a new version labeled as “ToN-IoT-R” in 
Table~\ref{tab:1}. Our proposed set of flow features is discussed in the following subsections. 

\begin{table}
  \caption{Dataset size and number of flow records in every attack category for different versions of ToN-IoT.}
  \label{tab:1}
  \centering
  \begin{tabular}{|c|c|c|c|c|c|}
    \hline
    Dataset & ToN-IoT  & ToN-IoT-R & NF-ToN-IoT-v2 & NF-ToN-IoT 
    & ToN-IoT  \\
    & (full) \cite{TNDp} & (this paper) & \cite{NFv2p} & \cite{NFp} & \cite{TNDp} \\
    \hline
    Number of & 45 & 12 & 45 & 14 & 45 \\
    flow features &  &  &  &  &  \\
    \hline
    Benign & 788'599 &	577'302 &	6'099'469  &	270'279 &	300'000  \\
    \hline
    Backdoor & 508'116	& 507'977 &	16'809 &	17'247 &	20'000  \\
    \hline
    DDoS & 6'165'008 &	5'525'554 &	2'026'234 & 326'345 & 20'000 \\
    \hline
    DoS & 3'375'328 &	3'258'564 &	712'609 &	17'717 &	20'000 \\
    \hline
    Injection & 452'659 &	328'256 &	684'465 & 	468'539 &	20'000  \\
    \hline
    MITM & 1'052 &	0 &	7'723 & 1'295	& 1'043  \\
    \hline
    Ransomware & 72'805 &	72'529 &	3'425 & 142 &	20'000  \\
    \hline
    Password & 1'365'958 &	1'307'215 &	1'153'323 &	156'299 &	20'000 \\
    \hline
    Scanning & 7'140'161 &	6'413'794 &	3'781'419  & 21'467 &	20'000 \\
    \hline
    XSS & 2'108'944	& 911'169 &	2'455'020  &	99'944 &	20'000 \\
    \hline
    Total Size & 21'978'630 & 18'902'360  & 16'940'496 & 1'379'274 & 461'043 \\
    \hline
  \end{tabular}
\end{table}

\subsection{Flow features}
\label{flow-feat}

The problem of identifying an optimal set of flow features that allows us to accurately detect various attacks is extremely 
important. It is also desired that special software should not be required to obtain such features. For instance, ToN-IoT feature 
set includes information about DNS, SSL, and HTTP activity which might not always be readily available in real world environment 
limiting its potential applications. On the contrary, a compact feature set proposed in \cite{NFp} relies solely on NetFlowv5 data making 
it extremely versatile. Unfortunately, by authors' own addmission it failed to provide sufficient attack detection accuracy.

NF-ToN-IoT-v2 is extended by adding derivatives of NF-ToN-IoT features such as numbers of packets with specific size, average values for certain 
categories making it as big as the original ToN-IoT feature set. Furthermore, it also includes features that encode connection 
flags (TCP, DNS, etc.) by enumerating them. This makes NF-ToN-IoT-v2 feature set more demanding to traffic monitoring software producing 
traffic samples for analysis. Furthermore, flow features are commonly normalized before GNN application \cite{TNDp, Egitp} leading to the 
information about enumerated flags being scrambled. This is aggravated further by the fact that when conventional statistical 
normalization techniques that rely on calculation of standard deviation and mean for a specific dataset are applied, 
flag information becomes impossible to restore. The latter operation also makes normalized features dataset-dependent.

To address both issues, we propose a new set of features that is more compact than NF-ToN-IoT-v2 feature set and can be derived from NetFlow data like 
NF-ToN-IoT features. We also propose normalization techniques that treat numeric features and flags differently. The proposed techniques are 
dataset-agnostic and allow us to preserve the intrinsic meaning of flag fields in the flows. 

\begin{table}
  \caption{The proposed flow feature set. Some flow features are converted into edge features 
  \textit{e\textsubscript{i}}} 
  \label{tab:2}
  \centering
  \begin{tabular}{|c|c|c|c|c|c|}
    \hline
    Flow feature & Edge feature   & Feature & Data type & Normalization & Normalization  \\
    number & number & & & method & coefficient \textit{k\textsubscript{w}}  \\
    \hline
    1 & - &  src\_ip &	string &  &  \\
    \hline
    2 & - &  src\_port &	uint16  &  &  \\
    \hline
    3 & - &  dst\_ip &	string  &  &  \\
    \hline
    4 & - &  dst\_port &	uint16  &  &  \\
    \hline
    5 & - &  protocol &	uint8  &  &  \\
    \hline
    \hline
    6 & 1 & duration &	numeric &	Eq. (\ref{eq:eni}) &	600  \\
    \hline
    7 & 2 &  src\_pkts &	numeric &	Eq. (\ref{eq:eni}) &	20  \\
    \hline
    8 & 3 &  src\_ip\_bytes &	numeric & Eq. (\ref{eq:eni}) &	900  \\
    \hline
    9 & 4 &  dst\_pkts &	numeric &	Eq. (\ref{eq:eni}) &	20  \\
    \hline
    10 & 5 & dst\_ip\_bytes &	numeric &	Eq. (\ref{eq:eni}) &	900 \\
    \hline
    - & 6 &  tcp &	binary &	- &	- \\
    \hline
    - & 7 &  udp &	binary &	- &	-  \\
    \hline
    - & 8 &  icmp &	binary &	- &	- \\
    \hline
    - & 9 &  duration=0 &	binary &	- &	-  \\
    \hline
    - & 10 & 1 & numeric &	- & -  \\
    \hline
    \hline
    11 & - &  Label &	binary  &  &  \\
    \hline
    12 & - &  Attack type &	string  &  &  \\
    \hline
  \end{tabular}
\end{table}

\subsection{The proposed feature set}
\label{proposed-feat}

In this paper we propose a new set of features and normalization techniques summarized in Table~\ref{tab:2}. 
We use six features of ToN-IoT dataset to obtain our feature set.  We use five numeric features as our features 6-10. 
We convert protocol information into binary flags corresponding to specific protocols (UDP, TCP, and ICMP). We also add a 
zero-duration flag that indicates single packet traverse or faulty connection. We do not use connection states since by analyzing 
header structure of TCP and UDP packets we conclude that sometimes UDP packet payload bytes may end up in TCP flags' data making 
this field unreliable. Finally, we add a unity feature which is later used for graph initialization in Section~\ref{GNN}.

As was previously discussed, using mean/std normalization for all fields leads to unfeasible results. Instead, we propose a 
dataset-agnostic normalization approach for numeric features. We use error function (erf) as normalization function for 
features 6 – 10:

\begin{equation}
  e_{ni} = \erf(\frac{e_{i}}{k_{wi}}),
  \label{eq:eni}
\end{equation}

where \textit{e\textsubscript{i}} is flow (edge) feature and \textit{k\textsubscript{wi}} is corresponding normalization coefficient 
shown in Table~\ref{tab:2}. Since our normalization 
function is a special case of sigmoid function, normalized feature values are in [0,1] range. Normalization coefficients 
roughly indicate half of the discrimination limit of the function, so coefficients are bigger for features with large 
variation range, e.g., for ip bytes and duration, and smaller otherwise, as in case with packets number. 

\section{Graph Neural Networks for network data analysis}
\label{GNN}

To analyze network flow data with GNNs Lo \textit{et al.} proposed E-GraphSage \cite{EGSp} algorithm which extended conventional GraphSage 
\cite{GSp} to account for edge features. This allows to construct a graph with nodes corresponding to ip:port pairs and 
edges corresponding to network flows. Therefore, flow features 5-10 along with several derived features in Table~\ref{tab:2} can be conveniently encoded as 
GNN edge features 1-10 (with protocol field converted into flags as discussed in Section~\ref{NIDS-datasets}), features 1-4 are used to produce 
the graph, and features 11 and 12 are used as edge labels. In our experiments we use E-GraphSage implementation \cite{Egitp} 
without residual connections proposed in that paper.  

E-GraphSage algorithm includes a neighborhood aggregation and feature encoding stages. For aggregation we use mean aggregator 
with at most 15 randomly selected neighbors to limit the computational time dependence of aggregation subroutine on neighborhood size 
\cite{GSp,Egitp}. For encoder we use a conventional linear layer with 64 neurons and ReLU activation. 
In addition, we modify the graph node initialization procedure. In \cite{EGSp} the authors 
propose to initialize graph nodes with unit features (vector of ones) because node features are claimed to be overwritten during 
training. However, by carefully inspecting Algorithm 1 \cite{EGSp} one may notice that the initial \textit{h\textsubscript{v}} values are concatenated with 
the learned ones, meaning that every node vector will carry the initial unity vector. To avoid this issue, we propose to 
initialize nodes with ten node features similar to edge features in Table~\ref{tab:2}. The first nine features are initialized with 
average values of connected edges which is equivalent to performing initial neighbor aggregation for all neighbors for edge 
features 1-9 in Table~\ref{tab:2}. For 10\textsuperscript{th} node feature we find a normalized sum of all corresponding 10\textsuperscript{th} edge features thus 
encoding node degree in one of the node features as

\begin{equation}
  h_{v10}^0 = \tanh(\ln(\frac{\sum_{i=1}^{N_{v}}e_{in10}}{k_{w10}})) = 
  \frac{N_{v}^2-k_{w10}^2}{N_{v}^2+k_{w10}^2}
  \label{eq:h}
\end{equation}

where \textit{h\textsuperscript{0}\textsubscript{v10}} is node feature 10 of node \textit{v} in GNN layer 0, 
\textit{e\textsubscript{n10}} is 10\textsuperscript{th} feature of edge \textit{i} in neighborhood of \textit{v} and \textit{N\textsubscript{v}} 
is the number of neighbors of node \textit{v}. Normalization coefficient \textit{k\textsubscript{w10}} is 1000 in our study. Adding this feature is important since fixed-size neighborhoods are evaluated according to \cite{Egitp,GSp}, 
so for large neighborhoods only small subsets are taken into account. By encoding node degree as a node feature, we 
indirectly provide GNN with information about the real number of neighbors which is crucial for detecting attacks like 
distributed denial of service (DDoS).

In GNN training process attackers IPs are commonly masked to prevent neural network from developing bias and marking all traffic 
from some IP as attacks \cite{TNDp}. This practice is implemented in E-GraphSage papers, too \cite{EGSp,Egitp}. However, the possibility 
to develop bias towards IPs exists mainly for methods that include IP and port data directly into features \cite{TNDp}. On the 
contrary, for graph methods ip:port data is needed only to construct the graph and it is not encoded as part of any feature 
vector. Furthermore, IP randomization can have unpredictable results on detection of attacks that rely on repeated flow 
patterns originating from the same machine. In our experiments we do not perform source IP randomization (masking) for any 
attack other than DDoS, which, on the contrary, is often performed using a distributed network of attacking machines so IP 
masking is realistic and allows to simulate vast attack community with limited testbed data.

\begin{table}
  \caption{Flow classification accuracy on ToN-IoT-R and previous versions of ToN-IoT dataset.}
  \label{tab:3}
  \centering
  \begin{tabular}{|c|c|c|c|c|c|c|c|}
    \hline
    Dataset & ToN-  & ToN- & ToN- & NF-ToN- & NF-ToN- & NF-ToN- & ToN- \\
    & IoT  & IoT & IoT & IoT & IoT & IoT-v2 & IoT-R \\
    & \cite{EGSp} & \cite{Egitp} & \cite{NFv2p}  & \cite{EGSp} & \cite{NFv2p} & \cite{NFv2p} & (ours) \\
    \hline
    Number of & 39 & 39 & 39 & 8 & 8 & 39 & 10 \\
    edge features &  &  &  &  &  &  &  \\
    \hline
    Benign & 0.91	& 0.99 &	0.94 &	0.92 &	0.99 &	0.99	& 0.98  \\
    \hline
    Backdoor & 0.08 &	0.96 &	0.31 &	0.99 &	0.98 &	1.00 &	0.99  \\
    \hline
    DDoS & 0.98 &	0.55 &	0.98 &	0.68 &	0.72 &	0.99 &	0.99 \\
    \hline
    DoS & 0.73 &	0.73 &	0.57 &	0.00 &	0.48 &	0.91 &	0.99 \\
    \hline
    Injection & 0.83 &	0.91 &	0.96	& 0.71 &	0.51 &	0.91 &	0.54  \\
    \hline
    MITM & 0.81 &	0.72 &	0.16 &	0.28 &	0.38 &	0.45 &	-  \\
    \hline
    Ransomware & 0.94 &	0.86 &	0.11 &	0.23 &	0.83 &	0.98 &	0.95  \\
    \hline
    Password & 0.91 &	0.99 &	0.92 &	0.25 &	0.24 &	0.97 &	0.86\\
    \hline 
    Scanning & 0.85 &	0.00 &	0.85 &	0.13 &	0.80 &	1.00 &	0.99 \\
    \hline
    XSS & 0.95 &	0.96 &	0.99 &	0.00 &	0.19 &	0.90 &	0.87 \\
    \hline
    Weighted Average & 0.87 &	0.94 &	0.87 &	0.63 &	0.60 &	0.98 &	0.97 \\
    \hline
  \end{tabular}
\end{table}

\section{Results and Discussions}
\label{Res-Disc}

\subsection{GNN prediction accuracy based on feature set and dataset version}
\label{GNN-acc}

Table~\ref{tab:3} compares our results with previously published ones. It shows that we achieve higher accuracy than NF-ToN-IoT almost reaching 
the NF-ToN-IoT-v2 level and even exceeding it in several categories while using only 10 edge features. Moreover, our proposed 
features coincide with NetFlowv5 data structure which is drastically different from NF-ToN-IoT-v2 approach. Since all flows 
corresponding to MITM have been filtered out during original traffic patterns cleanup (see Table~\ref{tab:1}) we do not provide 
accuracy results for that category. We attribute high accuracy in DoS and DDoS categories to the limited IP masking and 
graph initialization techniques proposed in this paper. The reason for low “injection” classification accuracy along with 
reduced accuracy in “password” and “XSS” categories will be investigated in the future.

\subsection{Discussions}
\label{Disc}

In Section~\ref{proposed-feat} we discussed that we preserve meaning of all flags in our feature set to allow additional analysis logic to be 
implemented. For instance, protocol flags (edge features 6-8 in Table~\ref{tab:2}) can be used to analyze UDP, TCP, and ICMP protocol 
flows independently. In this case each flow corresponding to a specific protocol can be passed to GNN that works exclusively with 
that protocol data. Since protocol data cannot realistically be mixed, this allows to obtain a more targeted GNN predictions.
GNN encoders are extremely compact, meaning that using several GNNs does not result in a sharp increase in computational 
burden. Unfortunately, ToN-IoT-R mostly consists of TCP attack flows, with UDP and ICMP flows being predominantly benign. 
This prevents us from showing experimental verification of the proposed approach in this paper and it will be presented in 
the future.

It should be stressed that only six flow features of the original ToN-IoT were used to derive ten edge features proposed in 
this paper. These features are also present in all publicly available NIDS datasets \cite{NFp}, meaning that the proposed approach 
can easily be implemented with those datasets, too. It is also remarkable that we achieve high accuracy without using DNS, 
SSL, and HTTP activity information. This indicates that this information is not useful for GNN, or rather GNN cannot learn 
useful information from it. This might be attributed to conventional enumeration and normalization techniques used in other 
studies that do not allow correct analysis of said features in neural network context. This indicates the importance of 
developing more suitable approaches for manipulating these data fields which should be different from the methods used for 
numeric data.

\section{Conclusions}
\label{conclusions}

This paper investigates different versions of ToN-IoT dataset and presents a new ToN-IoT-R dataset. In ToN-IoT-R several 
inconsistencies of the previous versions are eliminated by filtering out irrelevant traffic flow data and testbed artifacts, 
and relabeling some flows (e.g., normal DNS data exchange erroneously marked as attacks). Different feature sets are analyzed,
and new compact standardized feature set is proposed which requires only the information obtainable from conventional traffic 
monitoring NetFlowv5 capable equipment. 

Numerical features and flags are treated differently in the proposed preprocessing methodology and only numerical features 
are normalized. Dataset-agnostic sigmoid normalization functions are proposed to improve generalization potential of the 
method. E-GraphSage algorithm with modified node initialization procedure is used for flow classification and high accuracy 
is achieved using only 10 edge features. The achieved accuracy exceeds that of ToN-IoT and NF-ToN-IoT and even exceeds 
NF-ToN-IoT-v2 in several categories. 

The proposed method is highly versatile and can be applied to other publicly available NIDS datasets. Because the proposed 
feature set does not require specialized traffic monitoring software for data collection and aggregation, the method is
particularly promising for real-time traffic analysis NIDS.   

\section*{Acknowledgement}
\label{acknowledgement}

Authors would like to thank Dr Igor Netay for fruitful discussions.

\section*{Data Statement}
\label{Data-Statement}

ToN-IoT-R dataset is available on request for academic research purposes.


\bibliographystyle{IEEEtran}
\bibliography{IEEEabrv,ms}

\end{document}